\documentclass[aps,prb,superscriptaddress, twocolumn]{revtex4-2}
\usepackage{amsmath}
\usepackage{amssymb}
\usepackage{bm}
\usepackage{color}
\usepackage{appendix}
\usepackage{graphicx}
\usepackage{braket}
\usepackage{listings}
\usepackage{physics}
\usepackage{siunitx}
\AtBeginDocument{\RenewCommandCopy\qty\SI}

\begin{document}
\title{Efficient creation of shallow NV$^-$ ensembles by high-angle ion implantation}
\author{Kento Sasaki}
\email{kento.sasaki@phys.s.u-tokyo.ac.jp}
\affiliation{Department of Physics, The University of Tokyo, 7-3-1 Hongo, Bunkyo-ku, Tokyo 113-0033, Japan}
\author{Hideyuki Watanabe}
\affiliation{Global Research and Development Center for Business by Quantum-AI technology (G-QuAT), National Institute of Advanced Industrial Science and Technology, Tsukuba Central 2, 1-1-1 Umezono, Tsukuba, Ibaraki 305-8568, Japan}
\author{Tokuyuki Teraji}
\affiliation{Research Center for Electronic and Optical Materials, National Institute for Materials Science, 1-1 Namiki, Tsukuba, Ibaraki 305-0044, Japan}
\author{Takashi~Taniguchi}
\affiliation{Research Center for Materials Nanoarchitectonics, National Institute for Materials Science, 1-1 Namiki, Tsukuba, Ibaraki 305-0044, Japan}
\author{Kenji~Watanabe}
\affiliation{Research Center for Electronic and Optical Materials, National Institute for Materials Science, 1-1 Namiki, Tsukuba, Ibaraki 305-0044, Japan}
\author{Kensuke Kobayashi}
\email{kensuke@phys.s.u-tokyo.ac.jp}
\affiliation{Department of Physics, The University of Tokyo, 7-3-1 Hongo, Bunkyo-ku, Tokyo 113-0033, Japan}
\affiliation{Institute for Physics of Intelligence, The University of Tokyo, 7-3-1 Hongo, Bunkyo-ku, Tokyo 113-0033, Japan}
\affiliation{Trans-scale Quantum Science Institute, The University of Tokyo, 7-3-1 Hongo, Bunkyo-ku, Tokyo 113-0033, Japan}
\date{\today}

\begin{abstract}
Negatively charged nitrogen-vacancy (NV$^-$) centers located a few nanometers below the diamond surface are key quantum defects for nanoscale sensing of external spins.
However, the creation of shallow NV$^-$ centers with high yield remains a materials challenge.
Here, we demonstrate that high-angle ion implantation enhances the creation efficiency of shallow NV$^-$ centers.
By implanting $^{15}$N ions at angles exceeding $60^\circ$, we achieve high NV$^-$ yields approaching 10\% with effective NV$^-$ depths below 10~nm.
These yields are significantly higher than those typically reported for shallow NV$^-$ creation.
The enhanced NV$^-$ yield is consistent with an increased vacancy-to-nitrogen ratio in the near-surface region, which is expected to promote NV formation during annealing.
The created NV$^-$ ensembles show coherence properties comparable to those of single shallow NV$^-$ centers at similar depths, and allow detection of nuclear spins in van der Waals materials attached to the diamond surface.
Our results establish geometric control of ion implantation as a simple and broadly applicable approach to engineer shallow vacancy-related quantum defects.
\end{abstract}

\maketitle

\section{Introduction}

Negatively charged nitrogen-vacancy (NV$^-$) centers in diamond possess optically addressable and long-lived spins, and are widely recognized as a leading platform for quantum sensing~\cite{Rondin2014,Schirhagl2014,Degen2017}.
Owing to their atomic-scale size, NV$^-$ centers created at depths of only a few nanometers below the diamond surface enable the detection of nuclear spins in materials attached to the diamond surface~\cite{Mamin2013,Staudacher2013,Ohashi2013,DeVience2015,Ziem2019,Lovchinsky2017,Henshaw2022}. 
Such sensors capable of nanoscale nuclear magnetic resonance (NMR) and nuclear quadrupole resonance (NQR) measurements are scarce, highlighting NV$^-$ centers as a unique platform for nanoscale spin sensing~\cite{Budakian2024}.

In this research field, the creation of NV$^-$ centers with both shallow depth and high magnetic sensitivity is one of the key challenges. 
Charge instability of NV$^-$ centers in the near-surface region~\cite{Santori2009}, as well as noise originating from the surface and impurities~\cite{Rosskopf2014,Myers2014,Romach2015}, are known to limit the achievable depth and sensitivity. 
Stabilization of the NV$^-$ charge state requires appropriate surface treatments~\cite{Fu2010,Hauf2011,Kawai2019}. 
The physical cleanliness of the diamond surface contributes to enhanced quantum coherence~\cite{Sangtawesin2019,Dwyer2022}. 
Furthermore, a higher NV$^-$ yield, defined as the ratio [NV$^-$]/[N], is critical for achieving higher sensitivity, as nitrogen impurities act as paramagnetic noise sources~\cite{Bauch2020,Hayashi2020}.

A variety of approaches have been used to create shallow NV$^-$ centers to date. 
Low-energy nitrogen ion implantation is the most widely used technique~\cite{Pezzagna2010,Tetienne2018,Noda2025}. 
Nitrogen ions implanted at low acceleration energies (e.g., 2.5~keV) can introduce nitrogen atoms at depths of a few nanometers below the diamond surface. 
In such processes, a small implantation angle (typically $\sim7^\circ$) is commonly used to suppress ion channeling effects. 
Ion implantation masks~\cite{Ito2017,Ishizu2020,Speranza2025} and post-implantation etching~\cite{Loretz2014,FvarodeOliveira2015} have also been employed to achieve shallower nitrogen doping. 
However, the yield in these typical implantation-based techniques is known to be limited to approximately $0.1$--$1\%$, which is generally attributed to an insufficient number of vacancies created per implanted nitrogen atom, as well as charge instability in the near-surface region~\cite{Pezzagna2010,Rcke2021}. 
Although conversion efficiencies exceeding 5\% have been reported in experiments combining low-energy implantation of nitrogen or nitrogen-containing ions with thermal annealing in forming gas, such reports remain limited~\cite{Fukuda2018,Haruyama2019,Healey2026}.

Alternatively, shallow NV$^-$ centers can also be created by chemical vapor deposition (CVD) growth of nitrogen delta-doped diamond layers~\cite{Ohno2012,Ohashi2013,Ishiwata2017}. 
Nevertheless, the yield is typically as low as 0.3\% even in bulk diamond, resulting in values comparable to those obtained by low-energy ion implantation~\cite{Edmonds2012}. 
Attempts to create shallow NV$^-$ centers by introducing vacancies into nitrogen-doped diamond via ion implantation have also been reported, including nitrogen--helium co-implantation to enhance vacancy generation~\cite{NgandeuNgambou2022}.
However, the achieved conversion yields in related vacancy-engineering approaches have remained below 2.5\%~\cite{Healey2023}.

In this work, we demonstrate the efficient creation of shallow NV$^-$ ensembles using high-angle nitrogen ion implantation. 
By implanting $^{15}$N ions at implantation angles exceeding $60^\circ$ and applying appropriate thermal annealing, we demonstrate the formation of high-density shallow NV$^-$ ensembles with substantially enhanced yields.
From the signal strength of $^{11}$B spins in hexagonal boron nitride (hBN) flakes attached to the diamond surface, we estimate that the effective NV$^-$ depth is shallower than that of NV$^-$ centers created at comparable implantation energies with low implantation angles, and decreases systematically with increasing implantation angle.
This technique does not require specialized ion implantation systems, implantation masks, etching processes, or delta-doped growth layers, and therefore offers a simple, practical, and efficient route to the creation of shallow NV$^-$ centers.

\section{Main concept}

\begin{figure}
\begin{center}
\includegraphics{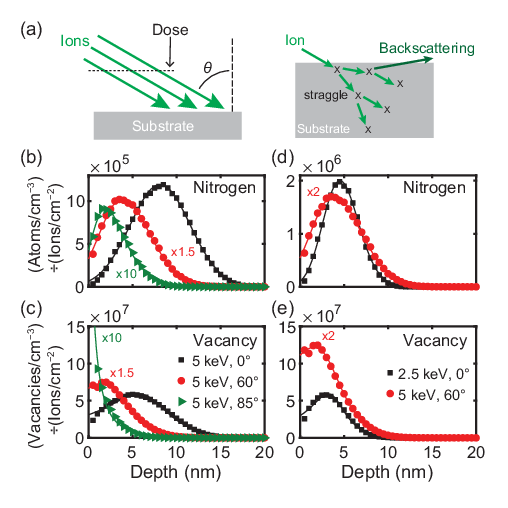}
\caption{
Introduction of nitrogen and vacancies by high-angle nitrogen ion implantation. 
(a) Schematic of high-angle ion implantation.  
The implanted ions undergo multiple scattering events within the lattice, resulting in straggle, and a fraction of the ions is backscattered (right panel). 
(b,c,d,e) Results of simulations of nitrogen ion implantation into diamond. 
(b,c) Depth distributions of implanted  (b) nitrogen and (c) vacancies at an implantation energy of 5~keV for different implantation angles. 
For clarity, the amplitudes of the distributions for $60^\circ$ and $85^\circ$ are scaled by factors of 1.5 and 10, respectively. 
(d,e) Comparison of the depth distributions of (d) nitrogen and (e) vacancies for conventional implantation and the present approach. 
For clarity, the amplitude of the distribution for the present approach (red circles) is scaled by a factor of 2.
The solid lines represent fits to the simulated profiles, using an asymmetric Gaussian for implanted nitrogen and an asymmetric Lorentzian for vacancies~\cite{Kehayias2021,Stancik2008}.
\label{fig1}
}\end{center}
\end{figure}

First, we describe the main concept of the present technique. 
This approach enables shallower introduction of nitrogen and vacancies, while simultaneously increasing the vacancy-to-nitrogen ratio, which is often insufficient for shallow NV$^-$ centers formation~\cite{Pezzagna2010,Rcke2021}. 
A schematic of this concept is shown in Fig.~\ref{fig1}(a). 
Nitrogen ions are implanted at a high implantation angle $\theta$ with respect to the diamond surface. 
As the implantation angle increases, the projected ion fluence onto the diamond surface is reduced by a factor of $\cos\theta$, reflecting the geometrical projection of the incident beam onto the surface (black dashed line in the left panel of Fig.~\ref{fig1}(a)).
The implanted ions undergo scattering inside the diamond, and oblique incidence causes them to stop at shallower depths compared to conventional normal-incidence implantation ($\theta\sim0^\circ$). 
Enhanced backscattering at high implantation angles leads to a further reduction in the retained nitrogen concentration relative to the nominal dose. 
As a result, this method increases the number of vacancies relative to nitrogen near the surface, which is expected to be favorable for shallow NV$^-$ centers formation.

We validate this concept using simulations based on the stopping and range of ions in matter (SRIM)~\cite{Ziegler2010_SRIM_first}. 
Figures~\ref{fig1}(b) and \ref{fig1}(c) show the calculated depth distributions of nitrogen atoms and vacancies, respectively, at an acceleration energy of 5~keV. 
By comparing the results for implantation angles of $\theta = 0^\circ$, $60^\circ$, and $85^\circ$, the simulations show that both distributions become shallower with increasing implantation angle. 
Due to enhanced backscattering at high implantation angles, the retained nitrogen density is reduced near the surface, whereas vacancies are still generated through collision events associated with backscattered ions.
Figures~\ref{fig1}(d) and \ref{fig1}(e) further compare the present approach ($\theta = 60^\circ$, 5~keV) with the conventional method ($\theta = 0^\circ$, 2.5~keV).
The present method provides a higher vacancy-to-nitrogen ratio near the surface while achieving nitrogen depth distributions comparable to those obtained by conventional low-angle implantation at lower acceleration energies (see Supplemental Material~\cite{SM} for more details).
In particular, this ratio becomes increasingly enhanced in the near-surface region, where vacancies are more readily annihilated during annealing.
These results indicate that high-angle implantation is effective for generating shallow nitrogen and vacancies.

\section{Experimental Methods}

To evaluate the effectiveness of the present technique, shallow NV$^-$ ensembles were created in $^{12}$C-enriched high-purity CVD grown layer~\cite{Watanabe1999} on (100)-oriented type-IIa diamond substrates using high-angle nitrogen ion implantation.
After cleaning, $^{15}$N$^+$ ions were implanted at an acceleration energy of 5~keV with a nominal dose of $1\times10^{13}$~cm$^{-2}$, corresponding to the dose set by the implantation system for normal incidence ($\theta=0^\circ$).
High implantation angles were achieved by tilting the diamond surface relative to the ion beam, resulting in effective implantation angles exceeding $60^\circ$.
Following implantation, the samples were thermally annealed in a forming-gas atmosphere to promote NV center formation, and subsequently oxygen-terminated by annealing in air to promote NV$^-$ charge stabilization.
Optical characterization of the resulting NV$^-$ ensembles was performed using a typical confocal microscope~\cite{Misonou2020}.
Further experimental details are provided in Ref.~\cite{SM}.

\section{Results and Discussion}

\begin{figure}
\begin{center}
\includegraphics{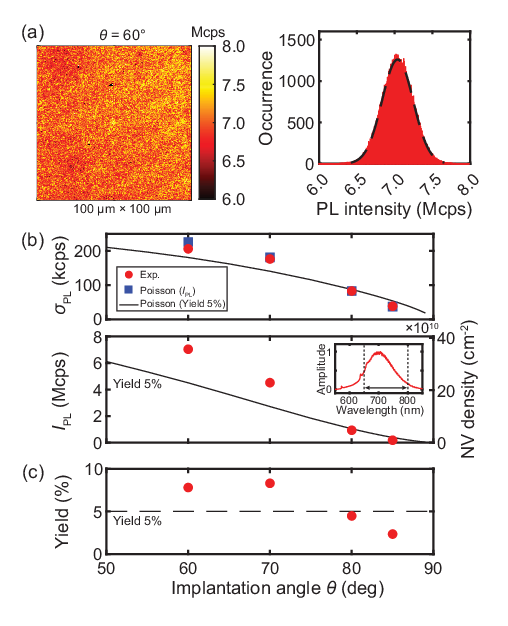}
\caption{
Photoluminescence (PL) intensity and its dependence on implantation angle $\theta$. 
(a) PL intensity on the surface of a sample implanted at $\theta = 60^\circ$. 
The left panel shows the PL intensity map acquired over an area of $(100~\mu\text{m})^2$ with a sampling resolution of 250~nm. 
The right panel shows the corresponding histogram of the PL intensity, displayed over a range of approximately $\pm5$ standard deviations of the distribution. 
(b) Dependence of the PL intensity on the implantation angle. 
The top and bottom panels show the standard deviation $\sigma_\text{PL}$ and average PL intensity  $I_\text{PL}$, respectively.
The inset shows a PL spectrum for the $\theta = 60^\circ$ sample; the dotted lines and arrows indicate the wavelength range used for PL detection. 
(c) Yield ([NV$^-$]/[N]) estimated from the average PL intensity. 
Red circles represent experimental results.
The black lines (crosses) indicate values corresponding to 5\% of the implanted nitrogen density estimated from SRIM simulations. 
Blue squares indicate the values estimated from the measured $I_\text{PL}$ under the assumption of a Poissonian spatial distribution.
\label{fig2}
}\end{center}
\end{figure}

We now present the experimental results obtained using the present technique. 
We first discuss the spatial distribution of the photoluminescence (PL) intensity. 
The left panel of Fig.~\ref{fig2}(a) shows the PL intensity map measured on the surface of a sample implanted at an implantation angle of $\theta = 60^\circ$. 
Except for a small number of localized spots exhibiting significantly different PL intensities, the PL intensity is spatially uniform over the entire area of $(100~\mu\text{m})^2$. 
The localized spots are likely caused by local surface topography that modifies the implantation conditions.

These PL signals predominantly originate from charge-stable NV$^-$ centers, as confirmed by the PL spectra discussed below.
The inset in the middle panel of Fig.~\ref{fig2}(b) shows a representative PL spectrum. 
The emission from neutrally charged NV centers (NV$^0$, zero-phonon line at 575~nm) is weak compared to that from NV$^-$ centers (zero-phonon line at 637~nm), and is almost negligible within the detection window (inside the black dashed lines).
Almost identical PL spectra are observed for samples implanted at all implantation angles (see Ref.~\cite{SM}).

The histogram of the PL intensity for all pixels is shown in the lower panel of Fig.~\ref{fig2}(a).
The histogram is well described by a normal distribution (black dashed line) with an average PL intensity $I_\text{PL} = 7.04$~Mcps and a standard deviation $\sigma_\text{PL} = 0.21$~Mcps.
Below we exclude pixels corresponding to anomalous bright or dark spots by analyzing only pixels with PL intensities within $\pm5$ standard deviations.
The top and bottom panels of Fig.~\ref{fig2}(b) show the dependence of $\sigma_\text{PL}$ and $I_\text{PL}$ on the implantation angle $\theta$, respectively. 
They decrease monotonically with increasing $\theta$. 
This trend can be attributed mainly to the reduction in the effective implantation dose with increasing $\theta$ (black dashed line in Fig.~\ref{fig1}(a)), together with changes in the NV$^-$ yield discussed below.

Using calibration data obtained from single NV$^-$ centers measured under identical conditions, we estimated the areal density of NV$^-$ centers from $I_\text{PL}$. 
The estimated NV$^-$ densities for implantation angles of $60^\circ$, $70^\circ$, $80^\circ$, and $85^\circ$ are $3.56\times10^{11}$, $2.29\times10^{11}$, $0.48\times10^{11}$, and $0.10\times10^{11}$~cm$^{-2}$, respectively (right axis in the bottom panel of Fig.~\ref{fig2}(b)). 
Dividing these values by the corresponding nitrogen density estimated from SRIM simulations ($4.57\times10^{12}$, $2.76\times10^{12}$, $1.06\times10^{12}$, and $0.40\times10^{12}$~cm$^{-2}$, respectively), we estimated yields of 7.8\%, 8.3\%, 4.5\%, and 2.4\%, respectively, as shown in Fig.~\ref{fig2}(c).
The black lines in Figs.~\ref{fig2}(b,c) represent a reference yield of 5\%.
These values are significantly higher than typical values for shallow NV$^-$ creation (0.1--1\%) and demonstrate the effectiveness of the present method.

The decrease in yield at high implantation angles is likely caused by the reduction in the effective nitrogen dose and enhanced backscattering, including reflection induced by the surface potential~\cite{Winter2002}.
These effects increase the nitrogen–vacancy separation compared to the typical surface–vacancy separation~\cite{Pezzagna2010,Rcke2021}.
Even under such high-angle conditions, higher yields are expected to be achievable by increasing the nitrogen implantation dose.
A more detailed understanding of the near-surface damage distribution will require further investigation using molecular dynamics simulations.

Alternatively, the NV$^-$ yield can also be estimated from the standard deviation $\sigma_\text{PL}$ by assuming a completely random (Poissonian) spatial distribution of NV$^-$ centers~\cite{SM}. 
Using this approach, the estimated yields for implantation angles of $60^\circ$, $70^\circ$, $80^\circ$, and $85^\circ$ are 6.5\%, 7.9\%, 4.5\%, and 2.5\%, respectively, which are again higher than those reported previously. 
These values are in good agreement with the NV$^-$ densities estimated from $I_\text{PL}$.
Similarly, the Poissonian estimates of $\sigma_\text{PL}$ derived from $I_\text{PL}$ are also consistent with the measured values (see blue squares in the top panel of Fig.~\ref{fig2}(b)).

To the best of our knowledge, only two previous studies have employed high-angle nitrogen ion implantation to create NV centers~\cite{Chakravarthi2021,Speranza2025}. 
In Ref.~\cite{Chakravarthi2021}, single NV$^-$ centers were created using implantation at $\theta=85^\circ$ and a high implantation energy of 85~keV, resulting in an estimated depth of $\sim20$~nm and a yield of about 1\%. 
Ref.~\cite{Speranza2025} combined nitrogen ion implantation at $\theta=45^\circ$ with a SiO$_2$ mask to localize nitrogen near the surface; however, the yield remained below 1\%. 
In contrast, the present work demonstrates the mask-free creation of shallow NV$^-$ ensembles with depths below 10~nm, as described later, and a high yield approaching 10\%, which are well suited for nanoscale NMR and NQR applications. 
While the total vacancy density is not expected to be dramatically higher than in conventional methods~\cite{SM}, the origin of the markedly enhanced yield in the present approach remains an open question and warrants further investigation.


\begin{figure}
\begin{center}
\includegraphics{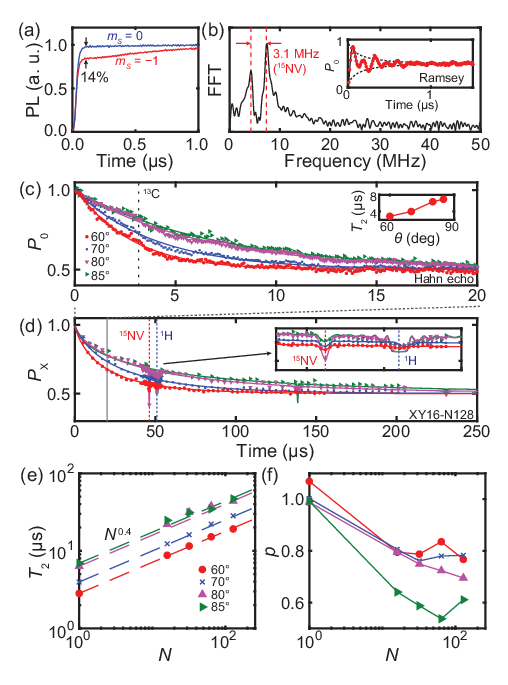}
\caption{
Characterization of the NV$^-$ spin properties. 
(a) Time-resolved PL measurement. 
The horizontal axis denotes the time after laser excitation.
(b) FFT amplitude of the Ramsey interference signal. 
The inset shows the time-domain data. 
In this measurement, a small magnetic-field misalignment of approximately $1.5^\circ$ was intentionally introduced to suppress the polarization of $^{15}$N. 
(c) Decoherence curve obtained with Hahn-echo sequence. 
The vertical black dotted lines indicate the expected signal position of $^{13}$C spins. 
The inset shows an implantation angle dependence of the coherence time.
(d) Decoherence curve obtained with XY16-N128. 
The vertical red and blue dotted lines indicate the NMR signals of $^{15}$N and $^{1}$H, respectively. 
The inset shows a magnified view around the NMR signals. 
(e) Dependence of the coherence time $T_2$ on the number of pulses $N$. 
(f) Dependence of the stretch exponent $p$ on the number of pulses $N$. 
All measurements were conducted at a magnetic field of approximately 29.5~mT, with a misalignment below $1^\circ$ relative to the NV axis, except for panel (b).
\label{fig3}
}\end{center}
\end{figure}

We next evaluate the fundamental sensing performance of the created NV$^-$ ensembles by characterizing their spin properties. 
Figure~\ref{fig3}(a) shows the time-resolved PL intensities for the $\theta = 85^\circ$ sample. 
A clear PL contrast between the $m_S = 0$ (blue) and $m_S = -1$ (red) spin states is observed for approximately 1~$\mu$s after laser excitation. 
The maximum contrast reaches about 14\%, and the contrast averaged over a 500~ns detection window is approximately 10\%. 
Although the contrast varies spatially, the average contrast ranges from about 5\% to 10\% for all implantation conditions, which is comparable to values typically observed for NV$^-$ ensembles in bulk diamond~\cite{Barry2020}. 
This high contrast is likely attributed to the suppressed contribution from undesired emitters such as NV$^0$.
In the following measurements, an initialization time of 5~$\mu$s was used to ensure sufficient spin polarization.

Then, we performed Ramsey interference measurements~\cite{Ramsey1950}. 
Figure~\ref{fig3}(b) shows a representative result obtained for the $\theta = 85^\circ$ sample. 
The Fourier-transformed spectrum exhibits a pair of clear peaks separated by 3.1~MHz, consistent with the hyperfine interaction in $^{15}$NV$^-$ centers~\cite{Felton2009}. 
This result confirms that the NV$^-$ ensembles are formed from the implanted $^{15}$N ions, while the contribution from residual nitrogen in the CVD layer and the diamond substrate is negligible.
By fitting the time-domain data (inset of Fig.~\ref{fig3}(b)) with a decaying oscillation described by a stretched exponential function, we obtain a dephasing time $T_2^* = 139$~ns and a stretch exponent $p = 0.6$. 
For other implantation angles of $60^\circ$, $70^\circ$, and $80^\circ$, similar values of $T_2^* = 184$--272~ns and $p = 0.5$--0.7 were obtained.
The deviation from a single-exponential decay ($p \approx 1$) suggests a distribution of $T_2^*$ values that differs from that expected for conventional NV ensembles~\cite{Bauch2020,Dobrovitski2008}, which is likely associated with variations in $T_2^*$ among NV$^-$ centers at different depths.
Although the obtained stretch exponent is close to the value of $2/3$ expected for decoherence induced by a two-dimensional ensemble of dipolar-coupled spins~\cite{Davis2023}, this mechanism is unlikely to be the dominant contribution in the present samples, since the measured $T_2^*$ values remain similar despite large variations in the nitrogen density.

We next evaluate the coherence using Hahn-echo sequence~\cite{Hahn1950}.
Figure~\ref{fig3}(c) shows the decoherence curves measured for different implantation angles. 
No electron spin echo envelope modulation due to $^{13}$C spin~\cite{Rowan1965, Childress2006} was observed in any of the samples, indicating that the NV$^-$ ensembles were formed exclusively within the $^{12}$C-enriched CVD layer. 
By fitting the data with a stretched exponential function, we obtained $(T_2, p) = (2.83~\mu\text{s}, 1.1)$, $(3.97~\mu\text{s}, 1.0)$, $(6.32~\mu\text{s}, 1.0)$, and $(7.20~\mu\text{s}, 0.9)$ for implantation angles of $60^\circ$, $70^\circ$, $80^\circ$, and $85^\circ$, respectively. 
The decay shape is approximately single exponential ($p=1$) in all samples. 
As the implantation angle increases, the coherence becomes longer (inset of Fig.~\ref{fig3}(c)). 
Although higher implantation angles are expected to produce shallower NV$^-$ centers that are more susceptible to surface noise, the effective nitrogen dose and implantation-induced damage are simultaneously reduced, which likely leads to the observed behavior. 

We compare the decoherence rates with the nitrogen areal densities.
When normalized to the density at $60^\circ$, the corresponding relative nitrogen areal densities are 100\%, 60\%, 23\%, and 8.8\%~\cite{SM}.
Taking the $\theta = 60^\circ$ sample as a reference, the relative decoherence rates $1/T_2$ are 100\%, 71\%, 45\%, and 39\%, respectively.
Except for the $\theta = 85^\circ$ sample, the decoherence rate scales approximately with the nitrogen areal density.
The increasing deviation at higher implantation angles likely reflects additional surface-related noise contributions that act as an effective offset.
Indeed, the long coherence times exceeding 5~$\mu$s observed for $\theta = 80^\circ$ and $85^\circ$ are comparable to those reported for single shallow NV$^-$ centers at depths of 5--6~nm~\cite{Sangtawesin2019}.

The obtained coherence times can also be compared with previous reports on dense near-surface NV$^-$ ensembles.
Ref.~\cite{Tetienne2018} investigated NV$^-$ ensembles created by conventional low-energy nitrogen implantation with doses of $5\times10^{11}$--$10^{13}$~cm$^{-2}$ and reported $T_2=10$--12~$\mu$s for the lowest dose and $T_2=2.5$--3~$\mu$s for the highest dose, with NV depths of 10--15~nm for 4--6~keV implantation.
Ref.~\cite{DeVience2015} reported $T_2\sim3~\mu$s for an NV$^-$ ensemble with an areal density of $3.5\times10^{11}$~cm$^{-2}$ created by 6-keV $^{14}$N$^+$ implantation with a dose of $2\times10^{13}$~cm$^{-2}$, corresponding to a yield of 1.8\%.
The present coherence times are broadly consistent with the implantation-dose dependence reported in these studies, while our NV$^-$ ensembles are formed at shallower effective depths.
In addition to being more susceptible to surface-related noise at shallower depths, the effective nitrogen density per unit volume also increases as the NV layer becomes thinner.
Therefore, the present results suggest that our method maintains a reasonable level of spin coherence even under such near-surface, high-density conditions.

We note that vacancy-related defects created by the present process may also contribute to decoherence.
Such contributions could become non-negligible when the vacancy density is further increased relative to the density of nitrogen-related defects, for example by implantation of ion species other than nitrogen~\cite{SM}.

We further investigate the performance under dynamical decoupling, which is essential for nanoscale NMR and NQR applications. 
Figure~\ref{fig3}(d) shows decoherence curves measured using an XY16 pulse sequence with $N = 128$ pulses (XY16-N128)~\cite{Carr1954, Meiboom1958, Gullion1990}. 
For all samples, the coherence is significantly extended compared to Hahn-echo measurements. 
In addition, the increased number of pulses enhances the sensitivity and spectral resolution around 1~MHz, enabling detection of the NMR signals from protons and the $^{15}$N of NV$^-$ centers (inset of Fig.~\ref{fig3}(d)). 
We therefore attribute the observed proton signal to hydrogen associated with surface termination or incorporated within the diamond lattice~\cite{Sasaki2020}, since no oil-immersion objective was used and no materials were intentionally attached to the diamond surface after oxygen termination.
Fitting the decoherence curve yields $(T_2, p) = (19.2~\mu\text{s}, 0.8)$, $(28.4~\mu\text{s}, 0.8)$, $(43.9~\mu\text{s}, 0.7)$, and $(54.8~\mu\text{s}, 0.6)$ for implantation angles of $60^\circ$, $70^\circ$, $80^\circ$, and $85^\circ$, respectively. 
As observed for Hahn echo, samples implanted at higher angles exhibit longer coherence times.

Figures~\ref{fig3}(e) and \ref{fig3}(f) show the dependence of the coherence time $T_2$ and stretch exponent $p$ on the number of pulses in the XY16 sequence. 
For all implantation angles, $T_2$ increases approximately as the 0.4 power of the pulse number ($N^{0.4}$). 
This scaling is weaker than that expected for noise dominated by bulk nitrogen impurities ($\sim N^{0.7}$)~\cite{deLange2010}, but is consistent with previous reports on shallow single NV$^-$ centers, where surface noise dominates ($N^{0.2}$--$N^{0.7}$)~\cite{Sangtawesin2019}. 
These results suggest that surface-related noise contributes significantly even for samples where Hahn-echo coherence correlates with nitrogen density, except for the $\theta = 85^\circ$ sample. 
The fast noise components that are less efficiently suppressed by dynamical decoupling may originate from implantation-induced defects correlated with nitrogen density, or from nitrogen-related impurities whose spin dynamics become sufficiently fast to act as broadband noise sources.

We also observe that the stretch exponent decreases with increasing pulse number (Fig.~\ref{fig3}(f)).
This behavior may partly originate from a depth-dependent enhancement of $T_2$, since NV$^-$ centers at different depths experience distinct local noise environments.
We note that the stretch exponent appears to approach a value close to $2/3$ at large pulse numbers, which is consistent with decoherence induced by a quasi-two-dimensional ensemble of dipolar-coupled spins~\cite{Davis2023}.
However, based on the quantitative analysis presented in Ref.~\cite{SM} using the obtained NV$^-$ densities, the expected coherence times limited by NV--NV interactions are significantly longer than the experimentally observed $T_2$ values.
Therefore, while the observed decay form is compatible with a two-dimensional dipolar model, NV--NV interactions cannot be the dominant mechanism limiting the coherence in the present samples.


\begin{figure}
\begin{center}
\includegraphics{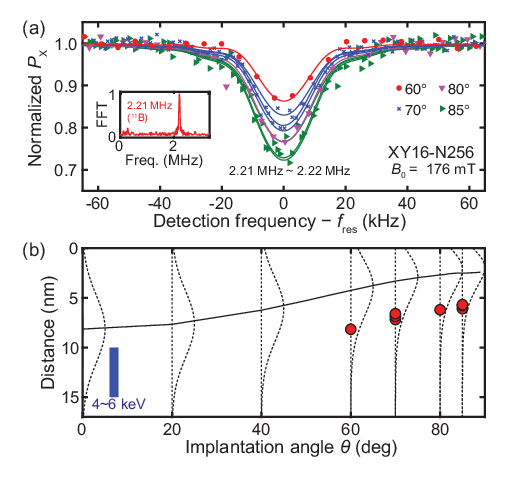}
\caption{
Depth estimation of NV$^-$ ensembles using detection of $^{11}$B spins in hBN flakes. 
(a) $^{11}$B spin spectra measured using an XY16-N256 sequence at a magnetic field of 176~mT. 
The transition probability $P_X$ on the vertical axis is normalized by a linear fit to the spectral edges to remove decay due to decoherence. 
The inset shows the spectrum obtained with the correlation spectroscopy. 
(b) Estimated distance (maximum effective depth) between the hBN flake and the NV$^-$ ensemble. 
Red circles represent experimental results. 
The black dotted line indicates the nitrogen depth distribution calculated by SRIM. 
The black solid line indicates the peak position of the nitrogen depth distribution estimated from the SRIM simulations.
Blue squares show the range of NV$^-$ depths obtained in Ref.~\cite{Tetienne2018}.
\label{fig4}
}\end{center}
\end{figure}

Finally, we evaluate the depth of the NV$^-$ ensembles created by the present method. 
Here, we directly apply a depth-estimation method originally developed for single NV$^-$ centers to NV ensembles and extract the maximum effective depth ($d_\text{nv}$)~\cite{SM}.
As discussed above, proton signals are detected even in the absence of intentionally attached materials, which requires caution when estimating NV depths using proton signals from an oil-immersion objective, as commonly employed in previous studies~\cite{Pham2016}. 
Instead, we attach hexagonal boron nitride (hBN) flakes to the diamond surface and detect the $^{11}$B spins~\cite{Lovchinsky2017,Henshaw2022}. 
This approach mitigates the risk of underestimating the NV depth, since unlike liquids, solid flakes do not conform to surface asperities and maintain a finite separation from the surface.
On the other hand, this method tends to overestimate the depth, as a hydrocarbon layer with a typical thickness of 1--2~nm is known to exist between hBN and the diamond surface~\cite{Lovchinsky2017,Henshaw2022}. 
Accordingly, we regard the NV--hBN separation estimated from the $^{11}$B signal as the $d_\text{nv}$.

Figure~\ref{fig4}(a) shows the detected $^{11}$B spin signals for all samples. 
Measurements were performed at one location for the $\theta = 60^\circ$ and $80^\circ$ samples, and at three locations for the $\theta = 70^\circ$ and $85^\circ$ samples. 
In all cases, a bias magnetic field of approximately 176~mT was applied along the NV axis, and an XY16-N256 was used. 
Signals were observed at frequencies consistent with the $^{11}$B resonance frequency $f_\mathrm{res}$ (2.21--2.22~MHz) for all samples. 
The inset of Fig.~\ref{fig4}(a) shows the Fourier spectrum of the correlation spectroscopy~\cite{Boss2016} acquired at this detection frequency, where a clear peak is observed at $f_\mathrm{res}$, supporting the assignment of the signal to $^{11}$B spins. 
Notably, the signal strength increases systematically with increasing implantation angle. 
This trend is consistent with a reduction in $d_\text{nv}$ with increasing implantation angle, in agreement with the central concept of the present work.

For quantitative analysis, we fitted the experimental spectra based on the theoretical models~\cite{Pham2016,Lovchinsky2017,Henshaw2022} to extract the $d_\text{nv}$. 
Details of the model are provided in Ref.~\cite{SM}. 
In the fitting procedure, the bias magnetic field, the NV--hBN separation ($d_\text{nv}$), and the nuclear-spin coherence time $T_{2,\mathrm{n}}^*$ were treated as free parameters, while the hBN flake was assumed to be sufficiently thick (100~nm). 
The solid lines in Fig.~\ref{fig4}(a) represent the fitted spectra, which show good agreement with the experimental data. 
The extracted nuclear-spin coherence times fall in the range $T_{2,\mathrm{n}}^* = 35.4$--66.7~$\mu$s, comparable to values reported in previous studies~\cite{Lovchinsky2017,Henshaw2022}.

Figure~\ref{fig4}(b) summarizes the implantation-angle dependence of the $d_\text{nv}$. 
Consistent with the nitrogen distributions predicted by SRIM simulations (dotted line), the distance decreases monotonically with increasing implantation angle. 
The extracted $d_\text{nv}$ (or their averages) for implantation angles of $60^\circ$, $70^\circ$, $80^\circ$, and $85^\circ$ were 8.2, 6.9, 6.2, and 5.9~nm, respectively.
From measurements at multiple locations for the $\theta = 70^\circ$ and $85^\circ$ samples, we estimate that the spatial variation in $d_\text{nv}$ is approximately 0.5~nm. 
This variation may originate from local differences in the NV--hBN contact condition, such as variations in the interfacial hydrocarbon layer thickness.

From the SRIM simulations, the mean depths and straggles for these conditions are estimated to be
$(4.8~\mathrm{nm}, \pm 2.5~\mathrm{nm})$, $(4.0~\mathrm{nm}, \pm 2.2~\mathrm{nm})$, $(3.3~\mathrm{nm}, \pm 2.0~\mathrm{nm})$, and $(3.2~\mathrm{nm}, \pm 1.8~\mathrm{nm})$, respectively.
The obtained $d_\text{nv}$ values are larger than the SRIM predictions by about 3~nm.
This offset exceeds that expected from a typical hydrocarbon interlayer alone and may originate from additional effects, such as residual ion channeling, which remain subjects for future investigation.

For comparison, we also plot representative values of $d_\text{nv}$ for NV$^-$ ensembles reported in Ref.~\cite{Tetienne2018} created at similar implantation energies (4--6~keV) using a conventional low implantation angle of $7^\circ$.
The data shown correspond to the range of 10--15~nm (blue squares in Fig.~\ref{fig4}(b)) estimated in Ref.~\cite{Tetienne2018} using proton NMR measurements~\cite{Pham2016}.
In contrast, the present method achieves shallower NV$^-$ ensembles ($d_\text{nv}<10$~nm) at comparable implantation energies, demonstrating the effectiveness of high-angle implantation for shallow NV$^-$ creation.

\section{Conclusion}

In conclusion, we have demonstrated the efficient creation of shallow NV$^-$ ensembles using high-angle ion implantation. 
By implanting $^{15}$N ions into high-purity diamond at implantation angles exceeding $60^\circ$, followed by thermal annealing, we achieved charge-stable NV$^-$ ensembles with a high yield approaching 10\%. 
Using detection of $^{11}$B spins in hBN flakes attached to the diamond surface, we showed that the maximum effective depth of the NV$^-$ ensembles can be systematically reduced with increasing implantation angle. 
The achieved depths ($d_\text{nv} < 10$~nm) are smaller than those typically reported for NV$^-$ ensembles created at comparable implantation energies using conventional low-angle implantation ($\gtrsim 10$~nm)~\cite{Tetienne2018}.

Toward further increasing the NV$^-$ density and deepening the understanding of this process, further optimization will be important.
Because the present method efficiently generates vacancies in the near-surface region, increasing the nitrogen dose may eventually enhance vacancy clustering or lead to amorphization of diamond.
For such high-density regimes, ion implantation under high-temperature conditions may help mitigate these effects by promoting vacancy diffusion during implantation.
A more comprehensive understanding of the distributions and dynamics of nitrogen atoms and vacancies, supported by molecular dynamics simulations, will also be important.

Beyond NV$^-$ centers in high-purity diamond, the present approach is expected to be applicable to the creation of vacancy-related quantum defects in a wide range of materials~\cite{SM}.
Examples include the formation of NV$^-$ centers in high-nitrogen, low-NV-density HPHT type-Ib diamond by implantation of species such as phosphorus~\cite{Healey2023}, as well as the creation of vacancy defects in other wide-bandgap materials such as hBN~\cite{Gottscholl2020} and SiC~\cite{Koehl2011,Falk2013,Soltamov2012}. 
The ability to efficiently create shallow vacancy-related quantum defects with minimal fabrication complexity makes the present technique a promising platform for quantum sensing and related applications.

\section{Acknowledgement}

We thank Taisuke Kageura (AIST) and Shinobu Onoda (QST) for fruitful discussions.
This work was partially supported by JST, CREST Grant No. JPMJCR23I2, Japan; Grants-in-Aid for Scientific Research (Nos. JP24K21194, JP25H01248, JP26H02007, JP25K00934, JP25K00931, and JP25K01292); 
New Challenge Research hosted by JSR Corporation via JSR-UTokyo Collaboration Hub, CURIE;
the Mitsubishi Foundation (Grant No. 202310021); the Cooperative Research Project of RIEC, Tohoku University;
``Advanced Research Infrastructure for Materials and Nanotechnology in Japan (ARIM)'' (No. JPMXP1225UT1155) of the Ministry of Education, Culture, Sports, Science and Technology of Japan (MEXT)
;``World Premier International Research Center Initiative on Materials Nanoarchitectonics (WPI-MANA)'' supported by MEXT;
MEXT Q-LEAP (JPMXS0118068379), JST Moonshot R\&D (JPMJMS2062), CSTI SIP ``Promoting the application of advanced quantum technology platforms to social issues.''


%

\end{document}